\documentclass[aps,pre,preprint,superscriptaddress,floatfix]{revtex4}

\usepackage{graphicx,latexsym}
\usepackage{bm}

\begin{document}


\title{The Effect of Patterned Slip on Micro and Nanofluidic Flows}


\author{S. C. Hendy}
\affiliation{Industrial Research Ltd, Lower Hutt, New Zealand}
\affiliation{MacDiarmid Institute for Advanced Materials
and Nanotechnology, School of Chemical and Physical Sciences,
Victoria University of Wellington, New Zealand}
\author{M. Jasperse}
\affiliation{MacDiarmid Institute for Advanced Materials
and Nanotechnology, School of Chemical and Physical Sciences,
Victoria University of Wellington, New Zealand}
\author{J. Burnell}
\affiliation{Industrial Research Ltd, Lower Hutt, New Zealand}



\date{\today}

\begin{abstract}
We consider the flow of a Newtonian fluid in a nano or microchannel with walls that have
patterned variations in slip length. We formulate a set of equations to describe the
effects on an incompressible Newtonian flow of small variations in slip, and solve these 
equations for slow flows. We test these equations using molecular dynamics simulations of 
flow between two walls which have patterned variations in wettability. Good qualitative agreement and 
a reasonable degree of quantitative agreement is found between the theory and the molecular dynamics 
simulations. The results of both analyses show that patterned wettability can be used
to induce complex variations in flow. Finally we discuss the implications of our results 
for the design of microfluidic mixers using slip.

\end{abstract}


\maketitle



\section{Introduction}

Several recent experiments \cite{Zhu01, Craig01, Zhu02, Craig03} report the measurement of large, shear-dependent liquid slip at partially wetting liquid-solid surfaces. While the origin of these dramatic violations of the no-slip boundary condition is still controversial \cite{Charlaix05}, interest is beginning to develop in how these effects may be exploited in microfludics \cite{Granick03}. Microfludics is undergoing rapid growth with applications to chemical and biochemical synthesis \cite{Sato03}, and high-throughput synthesis and screening \cite{Mitchell01}. These applications require the manipulation of fluids in microchannels where flows are limited to very low Reynolds numbers. As a result, mixing in microfluidic devices tends to be diffusion dominated, requiring long channels and long retention times to achieve good mixing. As the scale of this technology continues to diminish the effects of low Reynolds numbers will become more significant. However, the effect of slip at channel walls also increases at small length scales so it is natural to ask whether the effects of slip can be used to overcome some of the disadvantages of laminar flow \cite{Granick03}.

To increase mixing rates it is necessary to induce transverse or circulating flows in a
channel, increasing interfacial area between fluids or streamlines (for a recent review see \cite{Campbell04}). This can be achieved by active mixers, which possess moving parts, but these can be difficult to fabricate and maintain. Passive mixers on the other hand achieve mixing by virtue of their topology alone and have no moving parts. Suggested designs for passive mixers include using channels with patterned topography \cite{Stroock02,Stroock03}, or channels with patterned surface charge in electro-osmotic flows \cite{Erickson02}. Another possibility is to use chemically patterned channel surfaces. For example, Kuksenok and co-workers \cite{Kuksenok02,Kuksenok03a,Kuksenok03b} have modeled the mixing of a binary AB fluid in channels patterned with A-like and B-like regions.

Yet another approach might be to use patterned wettability to induce variations in slip. Slip is often characterised by a slip-length $\delta$, which is the distance at which the fluid velocity at a surface (i.e. the slip velocity) vanishes if it is linearly extrapolated beyond the surface. Measurements of slip lengths do vary widely but some groups have reported slip lengths of several microns \cite{Zhu01}. It is common to invoke the formation of nanobubbles at the hydrophobic surface \cite{galea04} to explain such large slip lengths. However there is still much disagreement about the magnitude of slip that can be induced although lengths of tens to hundreds of nanometers seem to be more typical \cite{Craig03,Zhu02}. Furthermore other factors such as surface roughness and surface contamination do seem influence slip length measurements considerably \cite{Craig03,galea04}. Lauga and Stone \cite{Lauga03} have recently considered the effect of patterned no-slip and no-shear stress regions in pressure-driven Stokes flow in a cylindrical geometry where the no-shear stress regions model the presence of micro or nanobubbles on channel walls. From this 
they derive an effective macroscopic slip-length which indeed is found to depend on shear-rate and geometry. 

Here we will assume that the variations in wettability can produce variations in the slip of the flow at the channel walls. Molecular dynamics simulations of flow past hydrophobic surfaces \cite{Barrat99a,Barrat99b,Barrat03} certainly demonstrate a strong relationship between wettability and slip, although the slip lengths found tend to be of the order of a few of tens molecular diameters at most. However, as noted above, the formation of nanobubbles at hydrophobic surfaces may well be able to induce very large slip lengths: thus large variations in wettability on a surface might be expected to produce large variations in slip length.

We begin by studying a Newtonian flow in a simple channel with a slip boundary condition
characterised by a slip-length $\delta$ that varies in space i.e.
$\delta = \delta(x)$. In the first instance we are interested in seeing the effect of a variable slip length on the flow, and secondly, in evaluating the possible exploitation of such effects in designing a microfluidic mixing device. As this is a preliminary study, we will approach the problem analytically, using a pertubative scheme to satisfy the slip boundary condition.

In section 4, we will use molecular dynamics simulations of the flow of a Lennard-Jones fluid
between two plates. The interaction between the plates and the fluid will be allowed to vary in
space in order to test the predictions of the analysis in sections 2-3. We conclude with a
discussion of the implications of our findings here for the design of mixers in microchannels
using chemical patterning.

\section{Equations for Flows with Spatially Varying Slip}

We start with the Navier-Stokes equations for a viscous incompressible fluid:
\begin{equation}
\label{momentum0}
\rho \left( \frac{\partial \bm{u}}{\partial t} + \bm{u} \cdot \bm{\nabla} \bm{u} \right) =
-\bm{\nabla} p + \mu \nabla^2 \bm{u},
\end{equation}
\begin{equation}
\label{mass0}
\nabla \cdot \bm{u} = 0,
\end{equation}
where $\bm{u}$ is the velocity field, $p$ the fluid pressure, $\rho$ is the fluid density
and $\mu$ is the fluid viscosity.

\begin{figure}
\resizebox{\columnwidth}{!}{\includegraphics{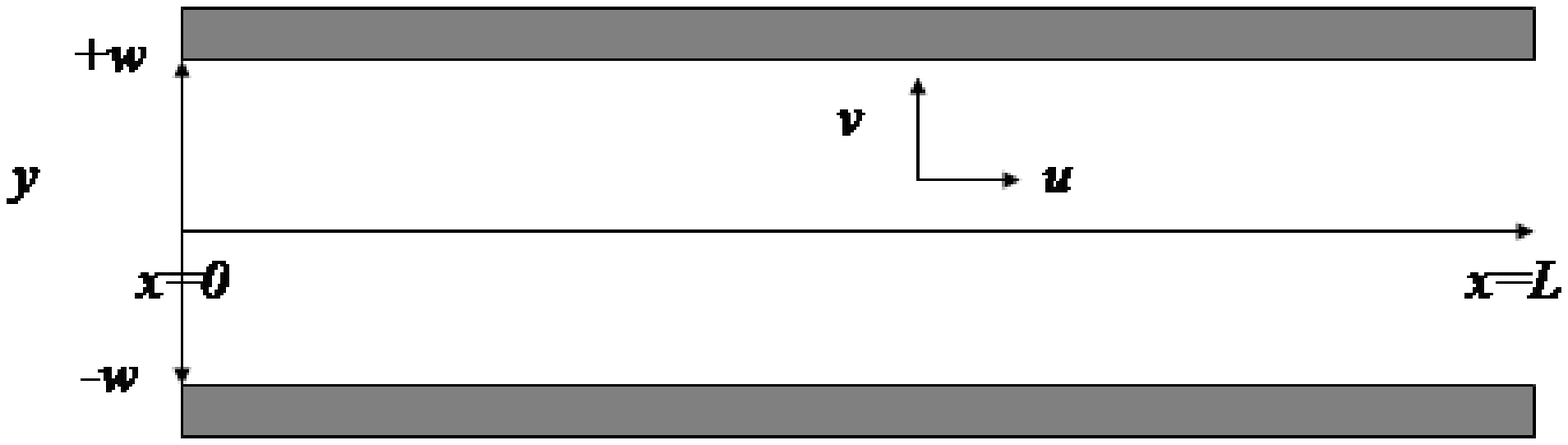}}
\caption{\label{geometry} The two-dimensional channel geometry.}
\end{figure}

We consider a pressure driven flow in a two-dimensional channel geometry corresponding to flow between two plates as illustrated in figure~\ref{geometry}. The channel has length $L$ and width $2w$. At the channel walls we have Navier slip boundary condition \cite{Navier}:
\begin{equation}
\label{SlipBC}
u(\pm w)= \mp \delta \frac{\partial u}{\partial y}(\pm w)
\end{equation}
where $u$ is the longtunidal velocity component ($x$-direction). The transverse velocity component $v$ ($y$-direction) 
satisifies $v(\pm w)=0$ at the walls. At the channel exit and entry we prescribe the pressure to be $p_0$ and $p_L$ respectively giving a pressure head across the channel of $\Delta p = p_0-p_L$. 

The solution to (\ref{momentum0}-\ref{SlipBC}) is
\begin{eqnarray}
\label{ConstantSlipu} u & = & \frac{\Delta p}{\mu L} \left( w^2+2 w \delta - y^2\right)
= U \left(1+2\frac{\delta}{w}-\frac{y^2}{w^2} \right)\\
\label{ConstantSlipv} v & = & 0 \\
\label{ConstantSlipp} p & = & p_{in}-\Delta p \left( \frac{x}{L} \right)
\end{eqnarray}
where $U = \frac{w^2 \Delta p }{\mu L}$ is the maximum fluid velocity in the absence of slip.

We will now allow the slip length to vary in the $x$ direction i.e.
\begin{equation}
\label{formalBC}
u(\pm w)= \mp \delta(x) \frac{\partial u}{\partial y}(\pm w).
\end{equation}
Specifically, we will consider the following slip boundary condition:
\begin{equation}
\label{fullBC}
u(\pm w) = \mp \delta \left(1+\alpha e^{ik x} \right) \frac{\partial u}{\partial y}(\pm w).
\end{equation}

If $\alpha \ll 1$ then we can apply a perturbative approach
\begin{eqnarray}
u & = & u_0 + \alpha u_1 + O \left( \alpha^2 \right) \\
v & = & v_0 + \alpha v_1 + ... \\
p & = & p_0 + \alpha p_1 + ...
\end{eqnarray}
where $(u_0,v_0,p_0)$ solve the constant slip-length boundary condition problem
(\ref{momentum0}-\ref{SlipBC}). The equations for the first-order corrections in $\alpha$
are then given by
\begin{eqnarray}
\label{momentum1x}
\rho \left( u_0 \frac{\partial u_1}{\partial x}
+ v_1 \frac{\partial u_0}{\partial y}\right)
& = & - \frac{\partial p_1}{\partial x} + \mu \nabla^2 u_1 \\
\label{momentum1y}
\rho u_0 \frac{\partial v_1}{\partial x}
& = & -\frac{\partial p_1}{\partial y} + \mu \nabla^2 v_1
\end{eqnarray}
and
\begin{equation}
\label{mass1}
\frac{\partial u_1}{\partial x} + \frac{\partial v_1}{\partial y} = 0
\end{equation}
with boundary condition
\begin{equation}
\label{FirstOrderBC}
u_1(\pm w)= \frac{2 w \delta \Delta p }{\mu L} e^{i k x}
\mp \delta \frac{\partial u_1}{\partial y}(\pm w) +  O \left( \alpha \right).
\end{equation}
The boundary condition immediately suggests the solution ansatz $u_1 = e^{i k x} f(y)$.
Inserting this into equation (\ref{mass1}), we find that
\begin{equation}
\label{h-def}
v_1= -ik e^{i k x} h(y)
\end{equation}
where $h^\prime (y) = f(y)$ and $h(0)=0$ since $v_1(0)=0$ by symmetry.

We can now eliminate $p_1$ from (\ref{momentum1x}) and (\ref{momentum1y}) to obtain the following
ordinary differential equation for $h(y)$:
\begin{widetext}
\begin{equation}
\label{diffh}
-\frac{d^4 h}{dy^4}+\left(\frac{i k u_0}{\nu}+2 k^2\right) \frac{d^2 h}{dy^2} -
k \left(k^3+\frac{i}{\nu}\left(u_0 k^2 + \frac{d^2 u_0}{dy^2} \right)\right)h= 0
\end{equation}
\end{widetext}
where $\nu = \mu/\rho$ is the specific viscosity. In terms of $h$ the boundary condition
(\ref{FirstOrderBC}) becomes
\begin{equation}
\label{hBC}
\frac{dh}{dy}(\pm w)=  2 U\frac{\delta}{w}
\mp \delta \frac{d^2 h}{dy^2}(\pm w)  + O\left( \alpha \right).
\end{equation}
We note that differential equation (\ref{diffh}) is homogeneous, so the magnitude of $h$ will be set by
the boundary condition (\ref{hBC}). Further when $\alpha \ll 1$, (\ref{diffh}) and
(\ref{hBC}) form a quasilinear boundary value problem. In the following section we will
examine the solution to this problem in a number of limiting cases.

\section{Relevance to microfluidic devices}

At this stage we will introduce some scales into the problem. As the effects of boundary slip
on the flow scale as $\delta / w$ (see equation~\ref{ConstantSlipu}), at widths substantially greater than
the slip length, a surface with patterned wettability will have an insignificant effect on the flow. With 
values of the slip-length of up to several microns reported in the literature \cite{Zhu01}, we will
confine ourselves to discussion of channels with widths $w < 10 \, \mu \mbox{m}$.

Obviously the wavelength of the slip variations $2 \pi/k$ is bounded above by the length of the
channel $L$. The wavelength is bounded below by the minimum size on which the slip can be
patterned. While microcontact printing \cite{Wilbur96} or photolithography of hydrophilic or hydrophobic
self-assembled monolayers might be limited to wavelengths greater than several hundred $\mu$m,
in principle, it is still of interest to consider the limit as $2 \pi/k \sim 10$ nm. This might
achieved using a combination of self-assembly by block copolymers and lithography for example
\cite{Nealy03}. Hence it is reasonable to consider patternings that satisfy 
$10^7 \mbox{nm} > 2 \pi/k > 10$ nm.

\subsection{Slow flows with fine patterning}

In the limit where $\frac{U}{k \nu} \ll 1$ and $\frac{U}{k \nu} \ll k^2 w^2$ then equation
(\ref{diffh}) reduces to
\begin{equation}
\label{diffh-lim2}
\frac{d^4 h}{dy^4}-2 k^2 \frac{d^2 h}{dy^2}+k^4 h = 0.
\end{equation}
For instance, in the case of water which is flowing at 0.01 $\mbox{ms}^{-1}$ in a
10 $\mu$m-width channel (i.e. 1 nanoliter per second) then $U/k \nu \ll w^2 k^2 \ll 1$
for $1/k \ll 20 \mu$m. Note that equation (\ref{diffh-lim2}) is real (whereas the equation 
(\ref{diffh}) is complex) so the variation in longitudinal flow velocity is in phase with the 
variations in slip length while the variations in transverse flow velocity are $90^o$ out 
of phase with the variations in slip length (recall equation (\ref{h-def})).

\begin{figure}
\resizebox{\columnwidth}{!}{\includegraphics{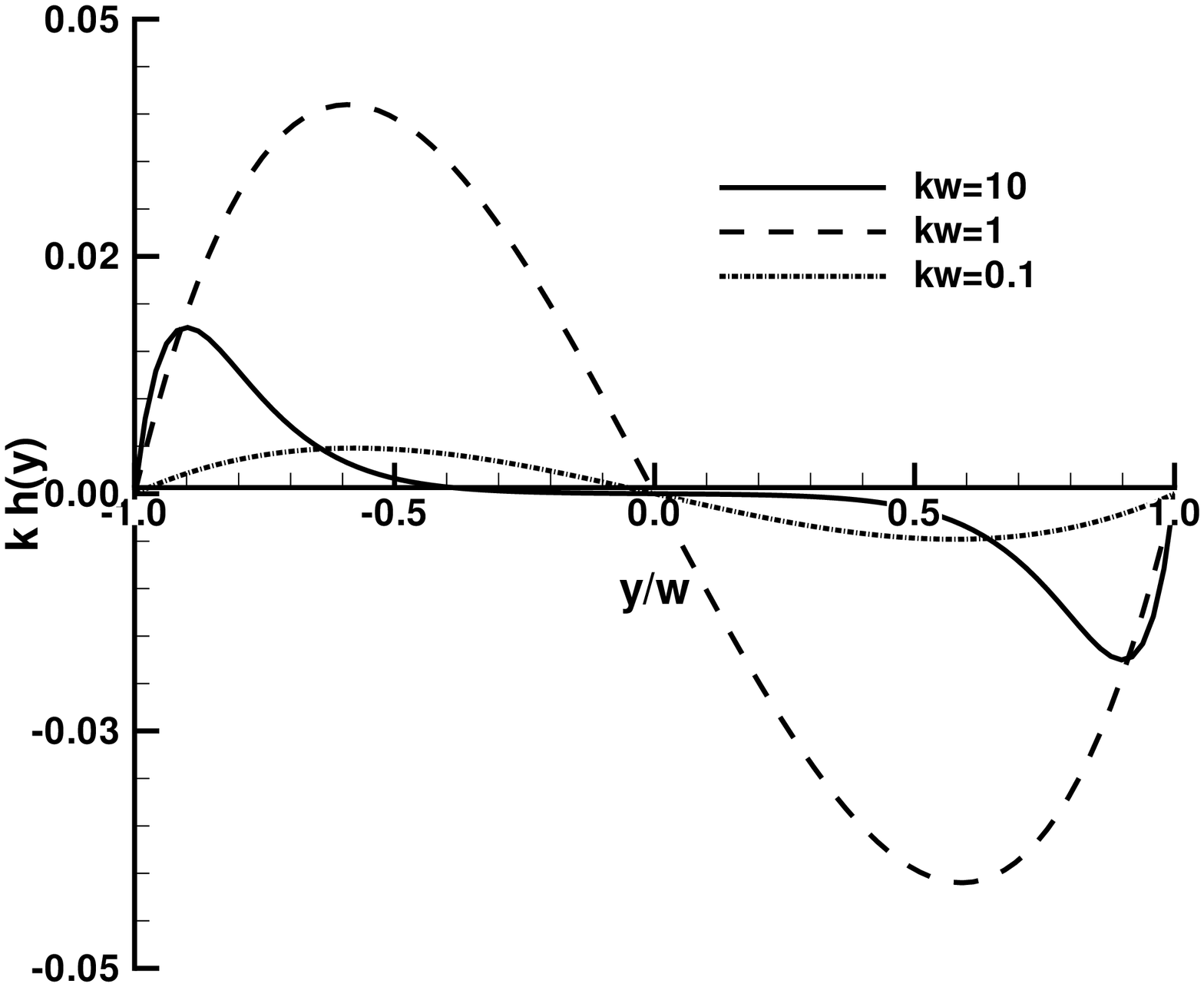}}
\caption{\label{h-plot} The function $k h(y) \sim v_1$ is shown in a channel for values of
kw=0.1,1,10 respectively. We have taken $\delta/w=1$.}
\end{figure}

\begin{figure}
\resizebox{\columnwidth}{!}{\includegraphics{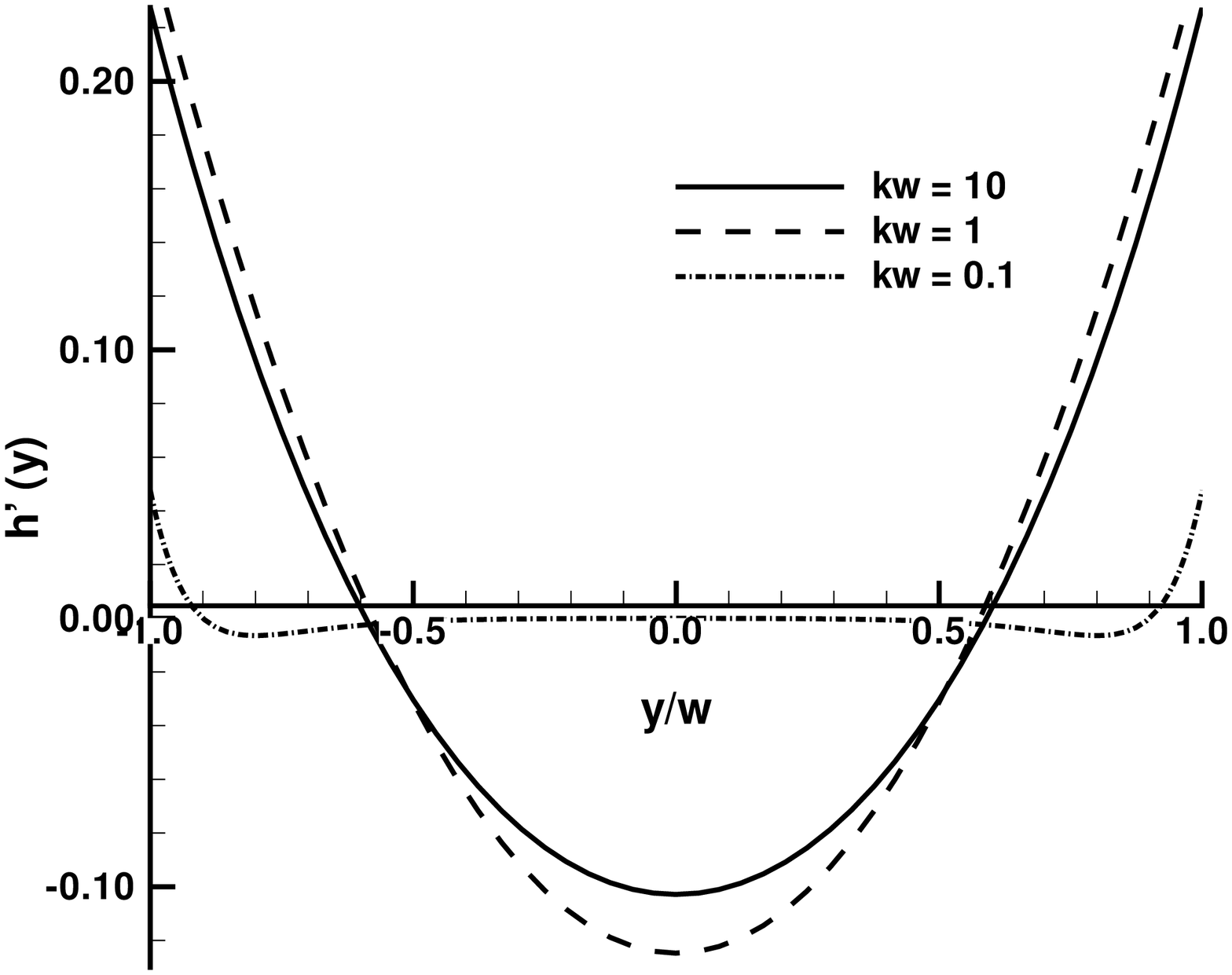}}
\caption{\label{dh-plot} The function $h'(y) \sim u_1 $ is shown in a channel for values of
kw=0.1,1,10 respectively. We have taken $\delta/w=1$.}
\end{figure}

The solution to (\ref{diffh-lim2}) with boundary condition (\ref{hBC}) to order $\alpha$ is 
given by
\begin{widetext}
\begin{equation}
\label{solh-1}
h \left( y \right) = U \left( \frac{\delta}{w} \right) \frac{(w-y) \sinh k \left(w+y \right)
-(w+y)\sinh k \left(w-y \right)}
{\sinh 2kw +2 k \delta \cosh 2kw - 4 k (w+\delta)}
\end{equation}
\end{widetext}
where we recall that $v_1=-i k e^{i k x}h(y)$ and $u_1=e^{i k x}h^\prime (y)$.
Figure~\ref{h-plot} shows $h(y)$ for $kw=$ 10, 1
and 0.1. It is clear from (\ref{diffh-lim2}) and figure~\ref{h-plot} that the magnitude of
$kw$ controls the variation away from the walls in $h(y)$ and hence in $v_1$ and $u_1$. With
$kw \ll 1$, then the transverse velocity induced $v_1$ is confined to very near the walls. 
Indeed, from figure~\ref{h-plot} we see that to maximize both the magnitude of $h(y)$, and 
its penetration towards the centre of the channel we should choose $kw \sim 1$. 
Similarly, figure~\ref{dh-plot} shows $h'(y)$ for $kw=10,1$ and 0.1. 

\begin{figure}
\resizebox{\columnwidth}{!}{\includegraphics{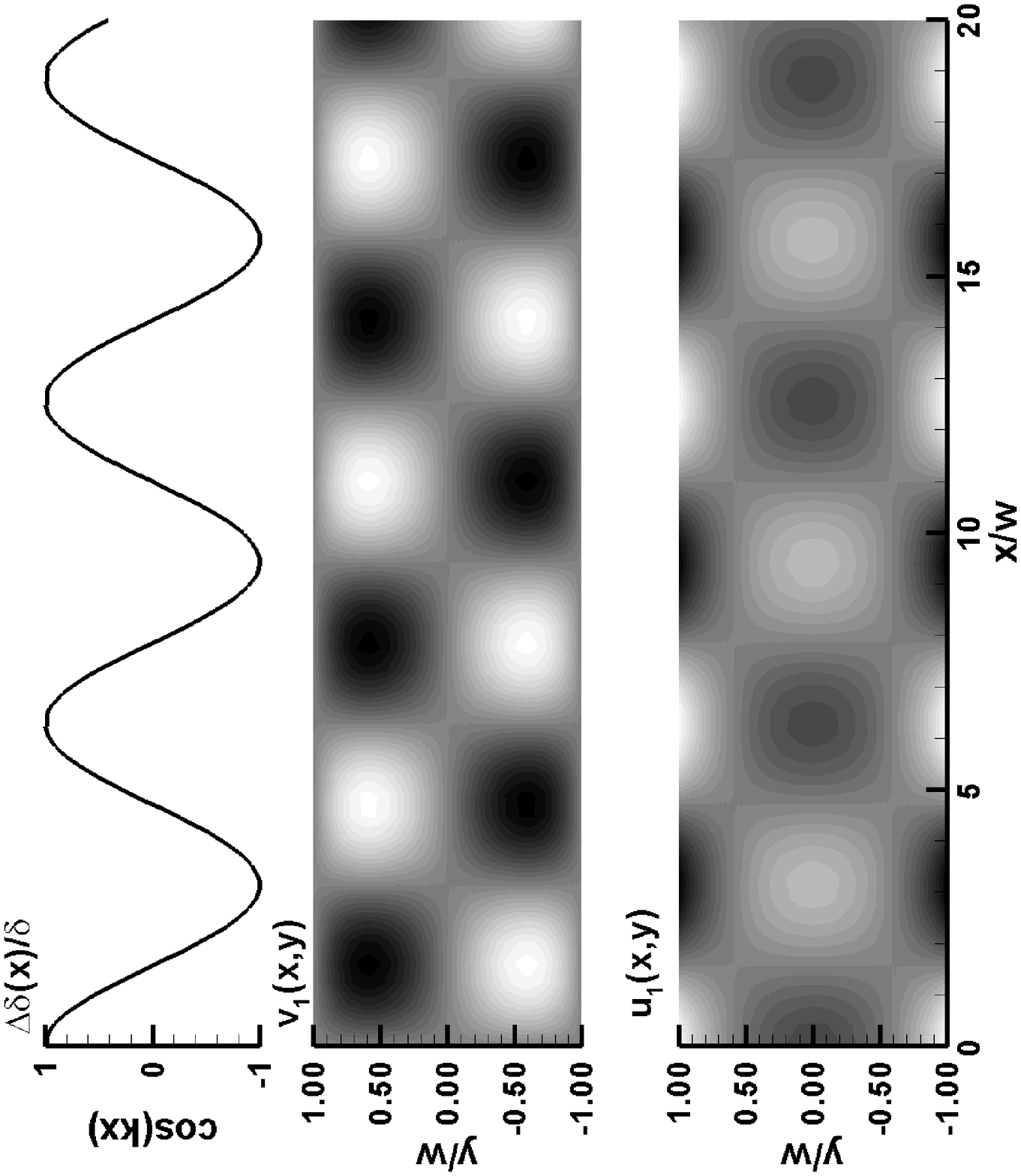}}
\caption{\label{v3d-plot} Contour plot showing $\Delta \delta(x) / \delta=\cos(kx)$ and the corresponding
variations in $v_1 (x,y)$ and $u_1 (x,y)$ in a channel for kw=1. Regions with dark shading
indicate negative velocity and regions with light shading indicate positive velocities.
We have taken $\delta/w=1$ and the channel length is $L=20 w$.}
\end{figure}

\begin{figure}
\resizebox{\columnwidth}{!}{\includegraphics{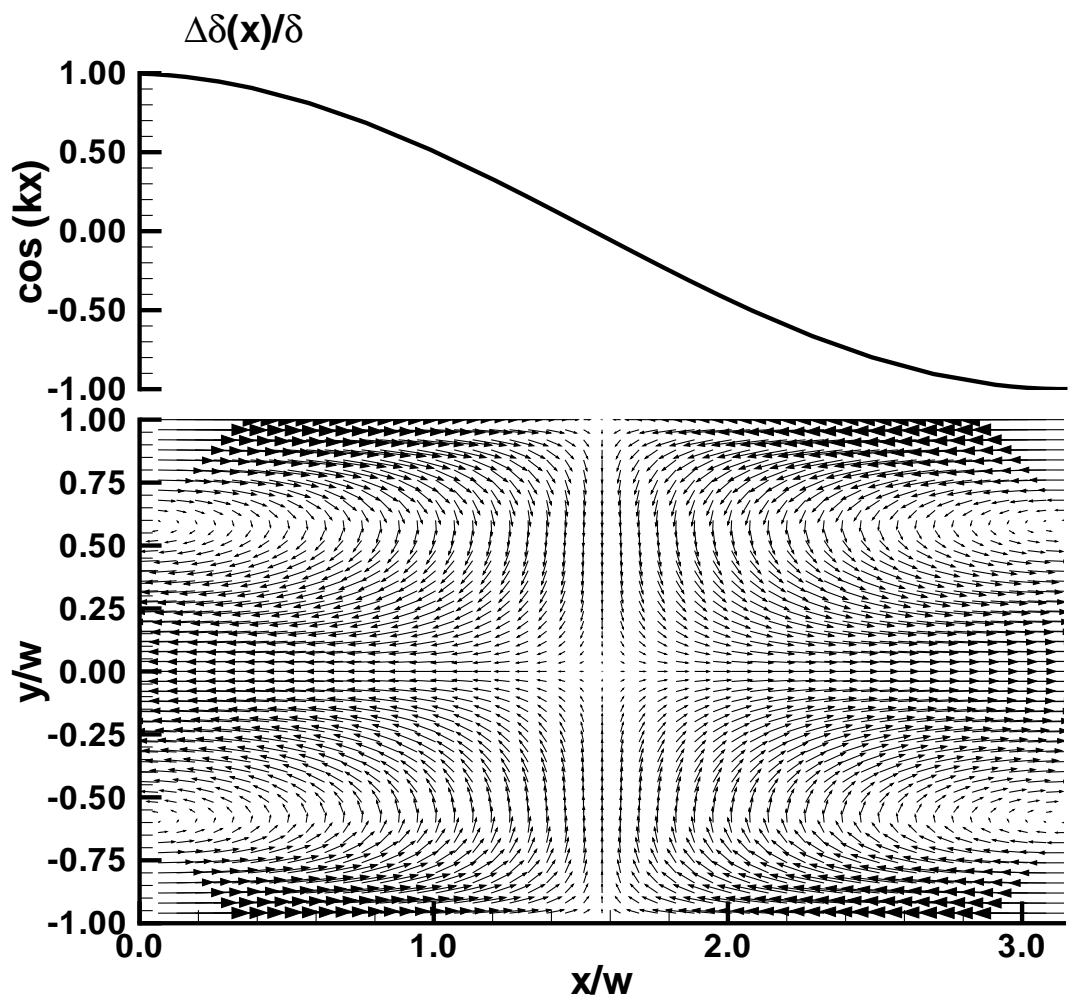}}
\caption{\label{vec-plot} Vector plot showing $\Delta \delta(x) / \delta =\cos(kx)$ and the corresponding
velocity vector $(u_1 (x,y), v_1 (x,y))$ in a channel for kw=1. We have taken $\delta/w=1$
and the channel length is $L=\pi w$.}
\end{figure}

Figure~\ref{v3d-plot} shows a flooded contour plot of the variations
in both velocity components ($u_1$ and $v_1$) along a long channel ($L=20 w$) with $kw = 1$,
and figure~\ref{vec-plot} shows a vector plot of the velocity components in a shorter channel
($L=\pi w$) with $kw = 1$. Note that where the slip at the channel wall is high, the flow velocity
increases at the channel walls, but {\em decreases} in the center of the channel. Likewise, where 
the slip is low, the flow velocity decreases at the channel wall but increases in the center of the 
channel. Between the peaks and troughs in slip, transverse flow is generated away from or towards 
the channel walls. 

\begin{figure}
\resizebox{\columnwidth}{!}{\includegraphics{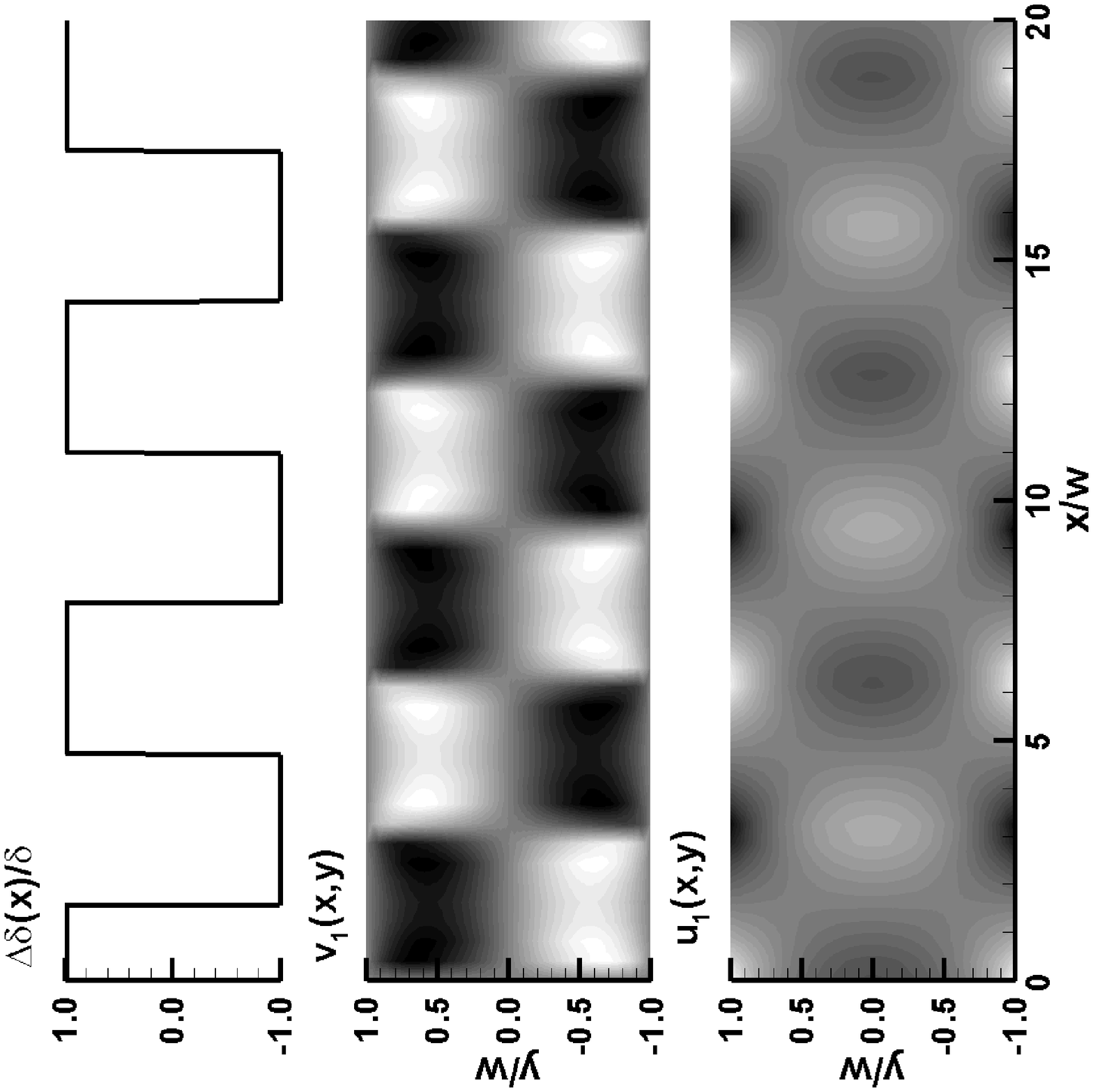}}
\caption{\label{v3dsq-plot} Contour plot showing a square wave $\Delta \delta(x) / \delta$ and the
corresponding variations in $v_1 (x,y)$ and $u_1 (x,y)$ in a channel for kw=1. Regions with
dark shading indicate negative velocity and regions with light shading indicate positive
velocities. We have taken $\delta/w=1$ and the channel length is $L=20 w$.}
\end{figure}

We can also look at square-wave variations in slip length, utilizing the Fourier series for
a square wave of wavelength $\lambda=2 \pi/k$:
\begin{equation}
f(x)=\frac{4}{\pi} \sum_{n=1}^\infty \frac{\sin((2n-1)kx)}{2n-1}.
\end{equation}
Since equation (\ref{diffh-lim2}) is linear we can solve for each Fourier mode and resum
to obtain the solution for a square wave variation in $\delta(x)$. Doing so gives
\begin{eqnarray}
u_1 & = & \frac{4}{\pi} \sum_{n=1}^\infty h'(k,y) \frac{\sin((2n-1)kx)}{2n-1} \\
\label{sq-v} v_1 & = & \frac{4}{\pi} \sum_{n=1}^\infty k h(k,y) \cos((2n-1)kx). 
\end{eqnarray}
Figure~\ref{v3dsq-plot} shows a flooded contour plot of the velocity components $v_1$ and
$u_1$ for a square wave variation in $\delta$ with $kw=1$.

\subsection{Slow flows with larger scale patterning}

\begin{figure}
\resizebox{\columnwidth}{!}{\includegraphics{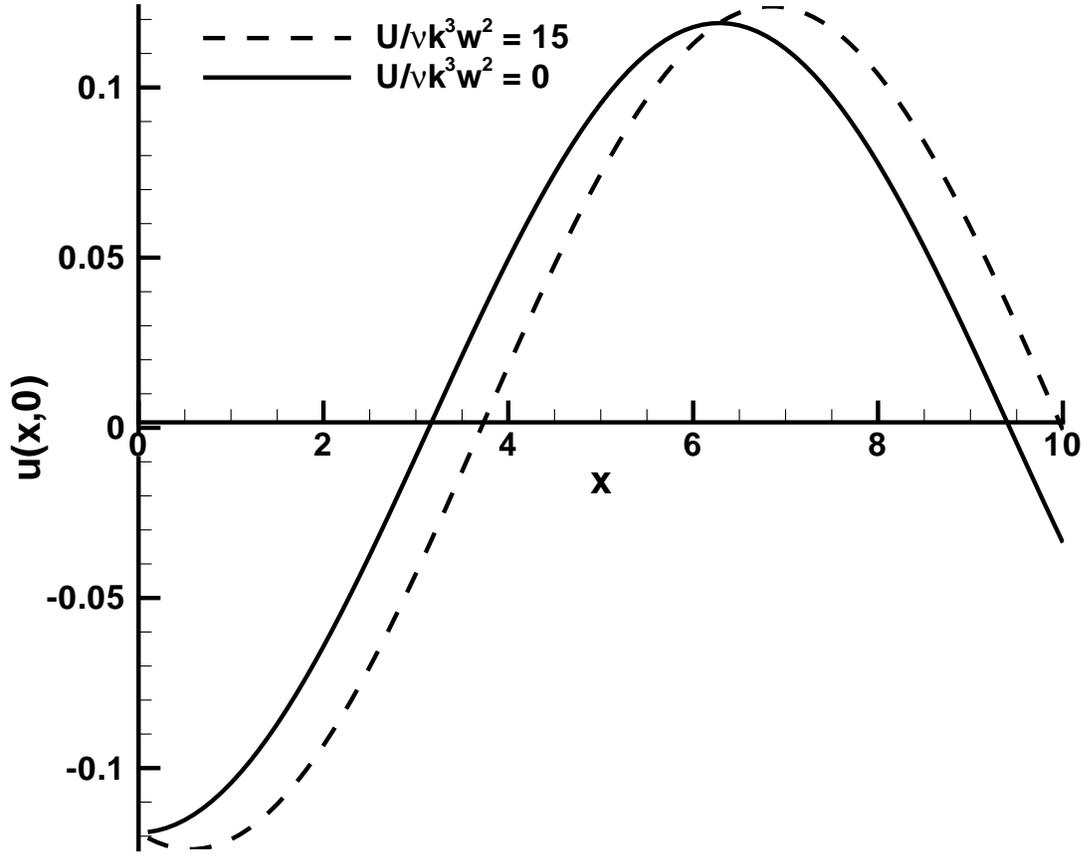}}
\caption{\label{ulag-plot} Plot of the change in velocity down the channel for
$\frac{U}{w^2 k^3 \nu}=15$ (solid line) and for
$\frac{U}{w^2 k^3 \nu}=0$ (dashed line) for kw = 0.1. We have taken $\delta/w=1$.
The effect of this term is to cause a phase lag in the velocity corrections away from
the walls (effectively shifting these changes downstream) and to increase the magnitude
of these corrections.}
\end{figure}

Now we consider the situation where $1 \gg \frac{U}{k \nu} \sim k^2 w^2$. For instance, in
the case of water which is flowing at 0.01 $\mbox{ms}^{-1}$ in a 10 $\mu$m-width channel
(i.e. 1 nanoliter per second) then $U/k \nu \sim w^2 k^2 \ll 1$ for $1/k \sim 20 \mu$m.
This corresponds to a slow flow with spatial variations in slip length occuring on scales
greater than the channel width. Now equation (\ref{diffh}) reduces to:
\begin{equation}
\label{diffh-lim3}
\frac{d^4 h}{dy^4}-2 k^2 \frac{d^2 h}{dy^2}+k\left( k^3+i\frac{U}{\nu w^2} \right) h = 0.
\end{equation}
The solution to this equation with boundary conditions (\ref{hBC}) is
\begin{widetext}
\begin{equation}
\label{solh-2}
h \left( y \right) =
\frac{U \left( \frac{\delta}{w} \right) \left(\sinh{\lambda_+ y} \sinh {\lambda_- w}
-\sinh{\lambda_- y} \sinh{\lambda_+ w} \right)}
{\lambda_+ \cosh{\lambda_+ w} \sinh{\lambda_- w} - \lambda_- \cosh{\lambda_- w} \sinh{\lambda_- w}
+\delta(\lambda_+^2-\lambda_-^2)\sinh{\lambda_+ w} \sinh {\lambda_- w}}
\end{equation}
\end{widetext}
where
\begin{equation}
\lambda_{\pm} = k^2 \sqrt{1\pm\sqrt{1-i\frac{U}{w^2 k^3 \nu}}}.
\end{equation}
Note that when $\frac{U}{w^2 k^3 \nu} \rightarrow 0$, the expression (\ref{solh-2}) for $h(y)$
reduces to the expression (\ref{solh-1}) from the previous section. In fact it is instructive
(although tedious) to write (\ref{solh-2}) in the form of (\ref{solh-1}) plus corrections in
$\frac{U}{w^2 k^3 \nu}$. Doing so we can write $h \left( y \right)$ as:
\begin{equation}
\label{phase-lag}
h \left( y \right) = h^{(1)} \left( y \right) + \frac{i U}{w^2 k^3 \nu} h^{(2)} \left(y\right)
+ O \left( \left( \frac{U}{w^2 k^3 \nu}\right)^2 \right)
\end{equation}
where $h^{(1)} \left( y \right)$ is given by equation~\ref{solh-1},
\begin{eqnarray}
h^{(2)} \left(y\right) & = & \frac{1}{12 \kappa} \left( 6 k (w^2-y^2) \sinh{ky} \sinh{kw}
+(w+y)(k^2(w+y)^2-3)\sinh{k(w-y)} \nonumber \right.\\
& & \left. -(w-y)(k^2(w-y)^2+3)\sinh{k(w+y)} \nonumber \right.\\
& & \left. + h^{(1)} \left( y \right) \left( 8k^3w^2 (3\delta +w)+ 6k^2 \cosh{2kw}
+3(4 \delta k^2w-1) \sinh{2kw} \right) \right)
\end{eqnarray}
and
\begin{equation}
\kappa = \sinh 2kw +2 k \delta \cosh 2kw - 4 k (w+\delta).
\end{equation}
Note that the first order correction in $\frac{U}{w^2 k^3 \nu}$ is purely imaginary. Thus it
introduces a phase lag the response of the fluid to the slip at the walls (moving it
downstream) and increases the magnitude of $h(y)$. This is shown in figure~\ref{ulag-plot}
which compares the velocity $u_1$ in the centre of the channel for $\Delta \delta = \delta \sin{kx}$ ($kw=0.1$) for $\frac{U}{w^2 k^3 \nu}=0$ and $\frac{U}{w^2 k^3 \nu} =15$ (a large value of $\frac{U}{w^2 k^3 \nu}$ is chosen here
so that the effect of this term is easily visible).

\section{Molecular dynamics simulations}

To study the effect of spatially varying wettability on flow in a channel at a molecular
level, we have followed the approach of Barrat and Bocquet \cite{Barrat99a,Barrat99b}. We consider a Lennard-Jones fluid with atomic mass $m$ confined between two walls with periodic boundary conditions applied in the plane of the walls. The walls consist of fixed Lennard-Jones atoms and interact with the fluid via a modified Lennard-Jones potential of the form:
\begin{equation}
\label{eq-lj}
\phi(r_{ij})=4\epsilon\left[\left(\frac{\sigma}{r_{ij}}\right)^{12}
-c_{fs}\left(\frac{\sigma}{r_{ij}}\right)^6\right]
\end{equation}
where $0 < c_{fs} \leq 1$ controls the degree of wettability of the walls \cite{Barrat99a}. 
Note that the fluid atoms also interact according to potential (\ref{eq-lj}) with
$c_{ff}=1.2$. Here we will consider flows where $c_{fs} =  c_{fs}(x)$ to model the effect
of chemical patterning of the channel walls.

We used a simulation cell containing 6750 fluid atoms within a volume of approximately $(20 \sigma)^3$. The
temperature was controlled using Nos\'{e}-Hoover thermostat \cite{Laird99} on the velocity component of the fluid atoms parallel to the channel walls but perpendicular to the imposed flow direction (in figure~\ref{geometry} this is the direction into the page). Flow can be induced by applying a body force to the fluid atoms in a direction parallel to walls giving a Poiseuille-type flow, or by dragging one of the walls past the fluid which induces a Couette flow. Here we only consider the former as our intention is to make a comparison with the pressure-driven flows of the previous section.

When $c_{fs}=1.0$ everywhere the flows are well-approximated by solutions to the incompressible Navier-Stokes equations (\ref{momentum0}-\ref{mass0}) with a no-slip boundary condition, although density variations occur near the 
walls due to the well-known tendancy for fluid atoms to layer at a solid interface. Furthermore, when $0.5 < c_{fs} < 1.0$ but is constant everywhere, we find that the flow is reasonably well-approximated by solutions to the Navier-Stokes equations with a slip boundary condition (\ref{SlipBC}). Our simulations are in good agreement with Barrat and
Bocquet \cite{Barrat99a,Barrat99b}.

\begin{figure}
\resizebox{\columnwidth}{!}{\includegraphics{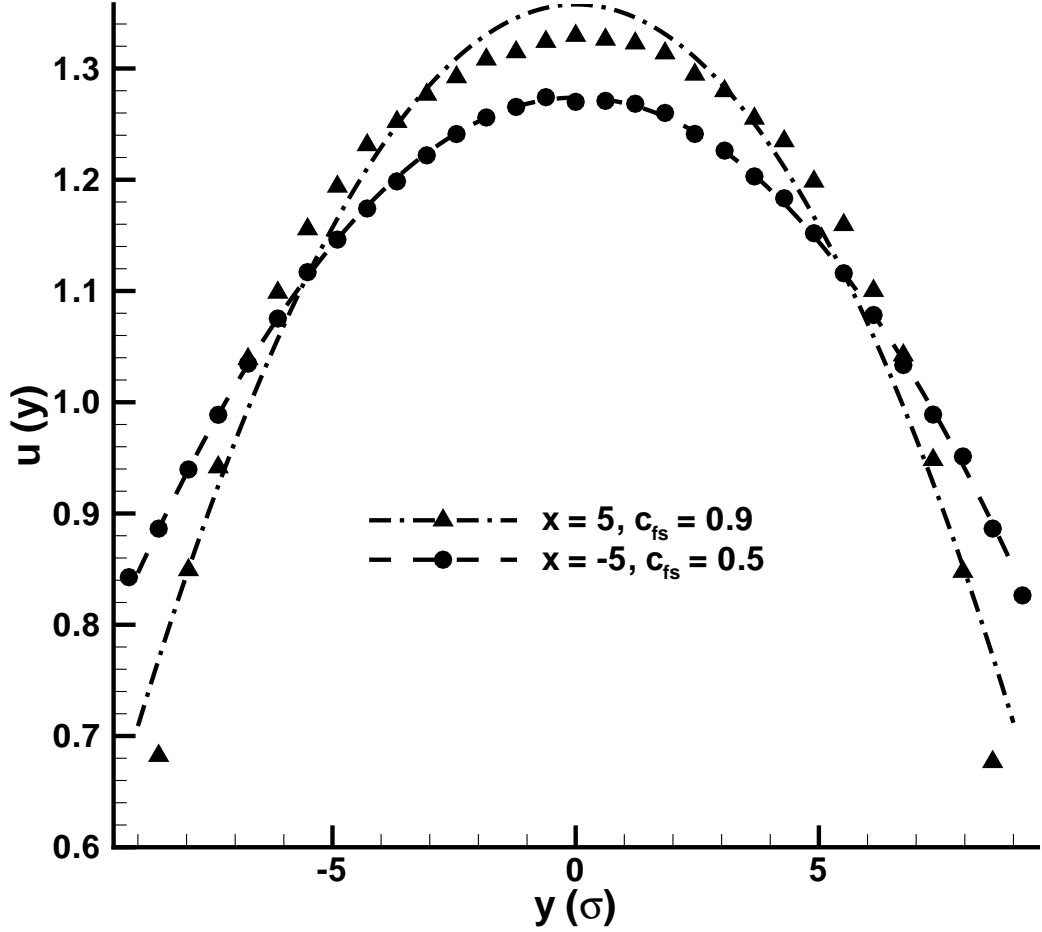}}
\caption{\label{patterned-slip} This plot shows the time-averaged longitudinal velocity
$u$ across the channel at $x=5\,\sigma$ (where $c_{fs} = 0.9$) and at $x=-5\,\sigma$
(where $c_{fs} = 0.5$). We calculate the effective slip length by fitting $U (1+\delta/w-y^2/w^2)$
to the profiles (fits are shown). In the solvophilic region ($x>0$ i.e. where $c_{fs}=0.9$) we 
calculated an effective slip length of $\delta = 9.1 \, \sigma$ and in the the solvophobic region ($x=-0.5$ $c_{fs}=0.5$) we calculated an effective slip length $\delta = 13.0 \, \sigma$.}
\end{figure}

We now consider the simulation of flows in a channel with patterned slip length. The equation
for $c_{fs}$ on the channel walls is:
\begin{equation}
c_{fs}= \left\{ \begin{array}{ll}
        0.9 & \,\,\, \mbox{if $\sin{kx} \geq 0$} \\
        0.5 & \,\,\, \mbox{if $\sin{kx} < 0$}
        \end{array} \right.
\end{equation}
where $kw=\pi$ (i.e. the wavelength of the pattern is 20 $\sigma$, which is the width of channel). Note that the mean value of $c_{fs}$ is 0.7. Our simulations show that such a patterning does indeed induce a variation in slip length along the channel walls. For instance, as illustrated in figure~\ref{patterned-slip} for a simulated flow with peak flow longitudinal flow velocity  $U=1.30\, (\epsilon/m)^{1/2}$, we calculated an effective slip length of $\delta = 9.1 \, \sigma$ by fitting a parabolic profile $U (1+\delta/w-y^2/w^2)$ to the longitudinal velocity profile in the solvophilic region ($x>0$ i.e. where $c_{fs}=0.9$). In the the solvophobic region ($x<0$ $c_{fs}=0.5$) we calculated an effective slip length $\delta = 13.0 \, \sigma$. Similarly, for a simulated flow with peak flow longitudinal flow velocity $U=0.4 \, (\epsilon/m)^{1/2}$, we calculated an effective slip length of $\delta = 3.6 \, \sigma$ in the solvophilic region ($x>0$ i.e. where $c_{fs}=0.9$). Likewise in the the solvophobic region ($x<0$ $c_{fs}=0.5$) we calculated an effective slip length $\delta = 6.7 \, \sigma$.

\begin{figure}
\resizebox{\columnwidth}{!}{\includegraphics{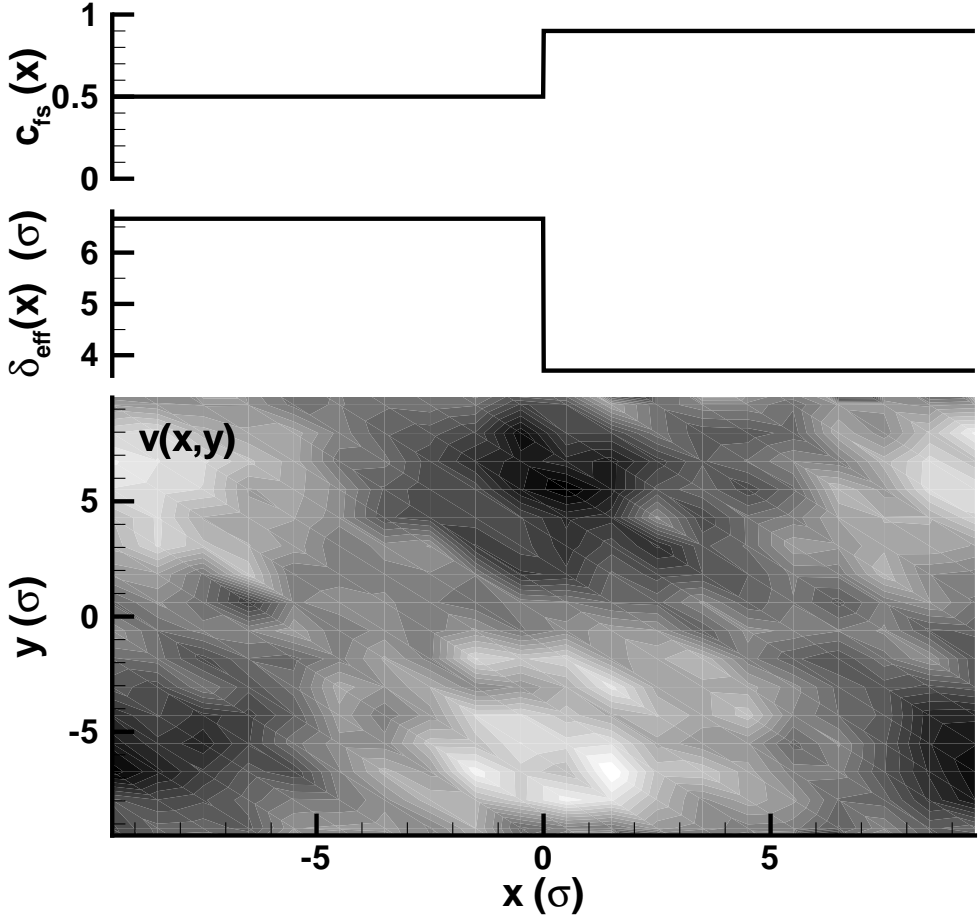}}
\caption{\label{MD-inphase} A plot showing the square wave $c_{fs}(x)$ boundary condition with
$kw=\pi$ imposed on the walls of the molecular dynamics simulation, the effective slip lengths
$\delta_{\mbox{eff}}$ induced by $c_{fs}$ and a corresonding contour plot showing
the variations in $v (x,y) = v_1 (x,y)$. The channel width is $2w=20 \sigma$ with
periodic boundary conditions applied at $x=\pm 10 \sigma$. The peak longitudinal flow velocity is $U=0.4 (\epsilon/m)^{1/2}$ and the peak transverse velocity is $V=0.03 (\epsilon/m)^{1/2}$.
Regions with light shading indicate flow in the $y$-direction and regions with light shading indicate flow in the negative $y$-direction. Note that the variations in $v(x,y)$ are $90^o$ out of phase with the variations in $c_{fs}$.}
\end{figure}

\begin{figure}
\resizebox{\columnwidth}{!}{\includegraphics{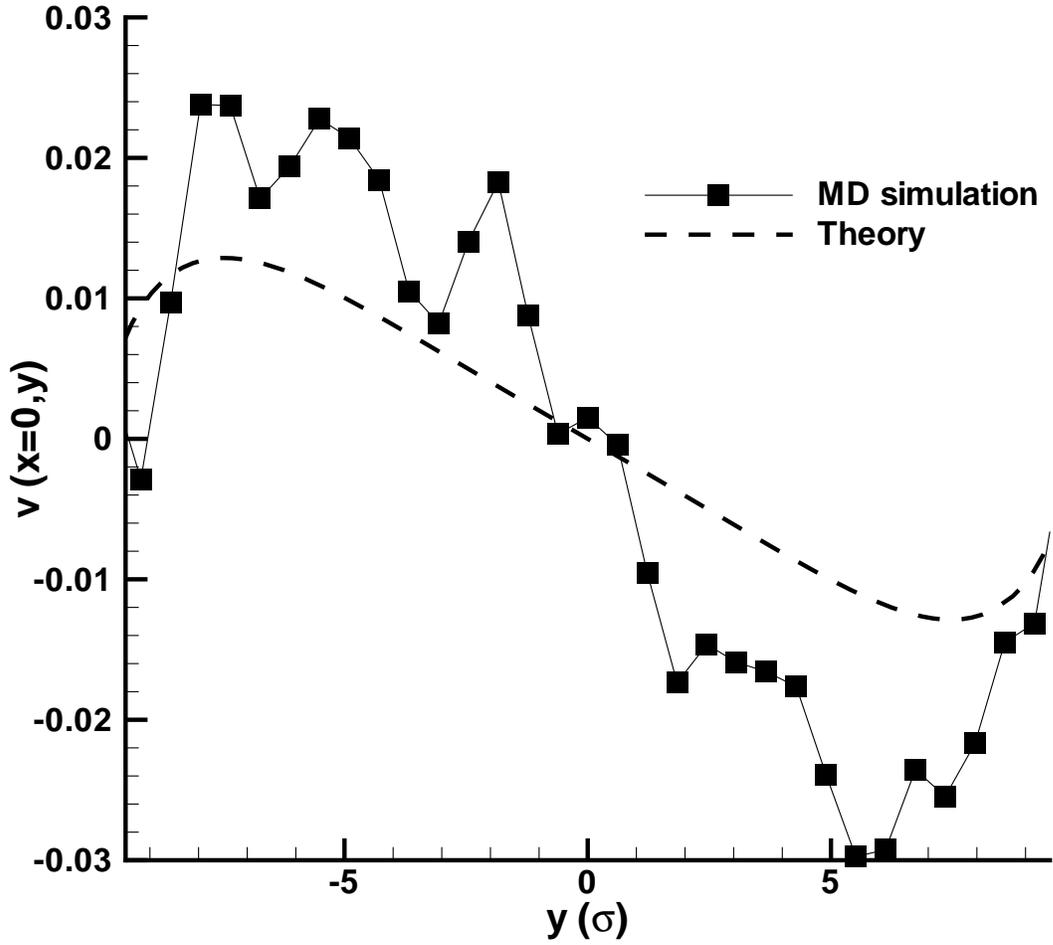}}
\caption{\label{MDvTh} A comparison of the transverse velocity $v$ at $x=0$ across the channel from the simulation in figure~\ref{MD-inphase} and from theory. The theory has been fitted 
to the simulation data by calculating an effective slip length $\delta = 6.7 \, \sigma$ across the solvophilic region ($x>0$ i.e. where $c_{fs}=0.9$) and an effective slip length $\delta = 3.6 \, \sigma$ across the solvophobic region ($x<0$ $c_{fs}=0.5$). Thus $\delta = 5.2 \, \sigma$ and $\alpha = 0.3$. It is seen from the comparison that the theory underestimates the peak values of $v$ by a factor of 2-3.}
\end{figure}

Furthermore, these variations in effective slip length induce transverse flows as anticipated in the previous sections. Figure~\ref{MD-inphase} shows the time-averaged transverse velocity of a flow, with peak longitudinal flow velocity $U=0.4 \, (\epsilon/m)^{1/2}$. The peak transverse velocity is $V=0.03 \, (\epsilon/m)^{1/2}$. Regions with light shading indicate flow in the $y$-direction and regions with dark shading indicate flow in the negative $y$-direction. Note that the variations in $v(x,y)$ are $90^o$ out of phase with the variations in $c_{fs}$ as predicted by our analysis in section 3. To compare the magnitude of the variations in $v$ from the simulation to the theory of the previous sections, we use the effective slip lengths calculated above. Thus $\delta = 5.2 \, \sigma$ and $\alpha = 0.3$
in equation (\ref{fullBC}). Figure~\ref{MDvTh} compares the theoretically expected variation in $v$ at $x=0$ across the channel for a square wave variation in slip length (see equation (\ref{sq-v})) to the time-averaged simulated variations. It is seen from the comparison that the theory underestimates the peak values of $v$ by a factor of 2-3.

\begin{figure}
\resizebox{\columnwidth}{!}{\includegraphics{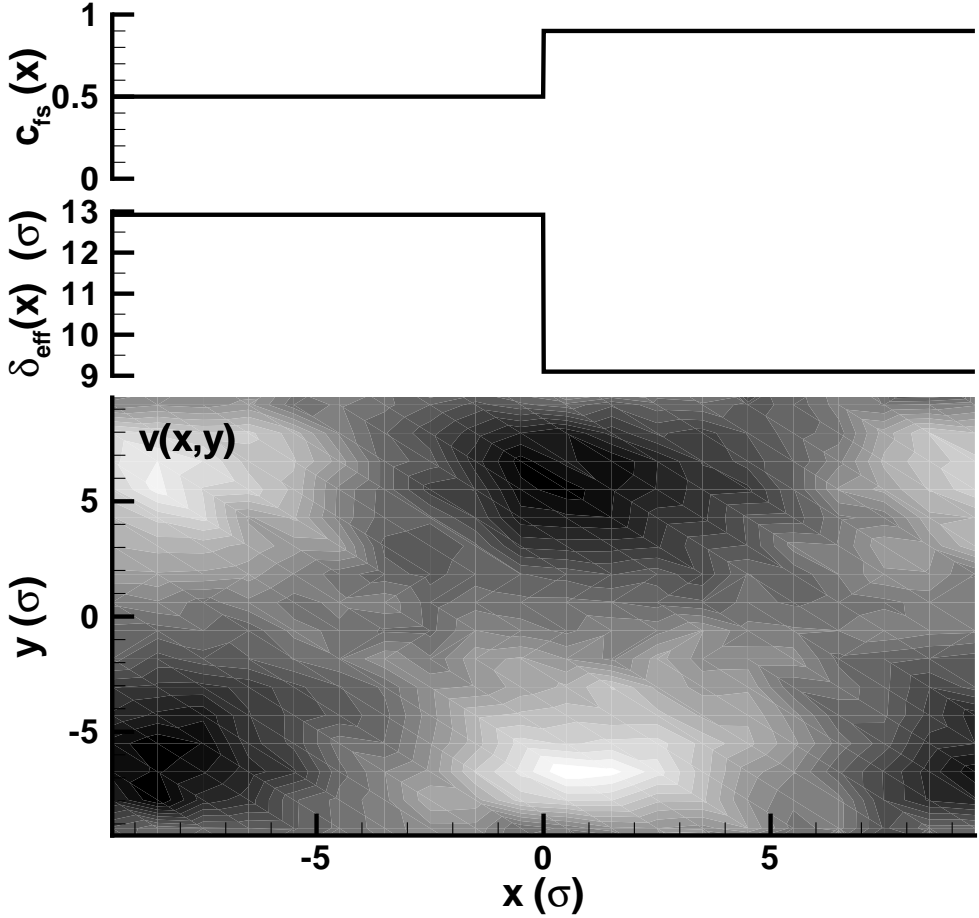}}
\caption{\label{MD-outofphase}  A plot showing the square wave $c_{fs}(x)$ boundary condition with $kw=\pi$ imposed on the walls of the molecular dynamics simulation, the effective slip lengths $\delta_{\mbox{eff}}$ induced by $c_{fs}$ and the corresponding
variations in $v (x,y) = v_1 (x,y)$ in a channel. The channel width is $2w=20 \sigma$ with periodic boundary conditions applied at $x=\pm 10 \sigma$. The peak flow velocity is
$U=1.30 (\epsilon/m)^{1/2}$ and the peak transverse velocity is $V=0.060 (\epsilon/m)^{1/2}$. Regions with light shading indicate flow in the $y$-direction and regions with light shading indicate flow in the negative $y$-direction.  Note the downstream phase shift in the variations in $v(x,y)$ with respect to the variations in $c_{fs}$, especially in comparion with figure~\ref{MD-inphase}.}
\end{figure}

Figure~\ref{MD-outofphase} shows a faster flow with peak longitudinal flow velocity is $U=1.30
\, (\epsilon/m)^{1/2}$ and peak transverse velocity is $V=0.060 \, (\epsilon/m)^{1/2}$. Regions
with light shading indicate flow in the $y$-direction and regions with dark shading indicate flow in the negative 
$y$-direction. Note the downstream phase lag in the variations in $v(x,y)$ with respect to the variations in $v(x,y)$ in the slower flow shown in figure~\ref{MD-inphase}. We have not made a direct comparison of this phase lag with the predicted phase lag in equation (\ref{phase-lag}) as we were unable to solve the full equation for $h$ (\ref{diffh}) for fast flows analytically. However, once again we find that the theory underestimates the peak values of $v$ by a factor of 2.

\section{Discussion}

The molecular dynamics simulations in section 4 demonstrate that patterned wettability will induce patterned variations in slip length. While this is not surprising given the demonstrated link between wettability and slip in other molecular dynamics simulations \cite{Barrat99a}, it supports the use of the boundary condition (\ref{formalBC}) in evaluating the effect of patterned wettability on flow. Furthermore we found a strong qualitative agreement between the molecular dynamics simulations and the approximate analytic solutions developed in sections 2-3, although the theory tended to underestimate the magnitude of the variations in flow due the patterning by a factor of 2. This underestimation may in part be due to the way the theory was fitted to the simulations (i.e. by fitting effective slip lengths in the different channel regions). However, the theory also assumes the fluid is incompressible, whereas substantial variations in fluid density can occur at the walls. In particular, a reduction in the density of the fluid near the solvophobic region of the wall relative to the solvophilic region of the wall, as observed in the MD simulations, would tend to enhance the transverse variations in flow. Finally, we note that in our analysis in sections 2-3 we only solved the slip boundary condition to order $\alpha$ (the relative variation in slip length). In the molecular dynamics simulations conducted here $\alpha$ was found to be 0.2-0.3.    
   
\begin{figure}
\resizebox{\columnwidth}{!}{\includegraphics{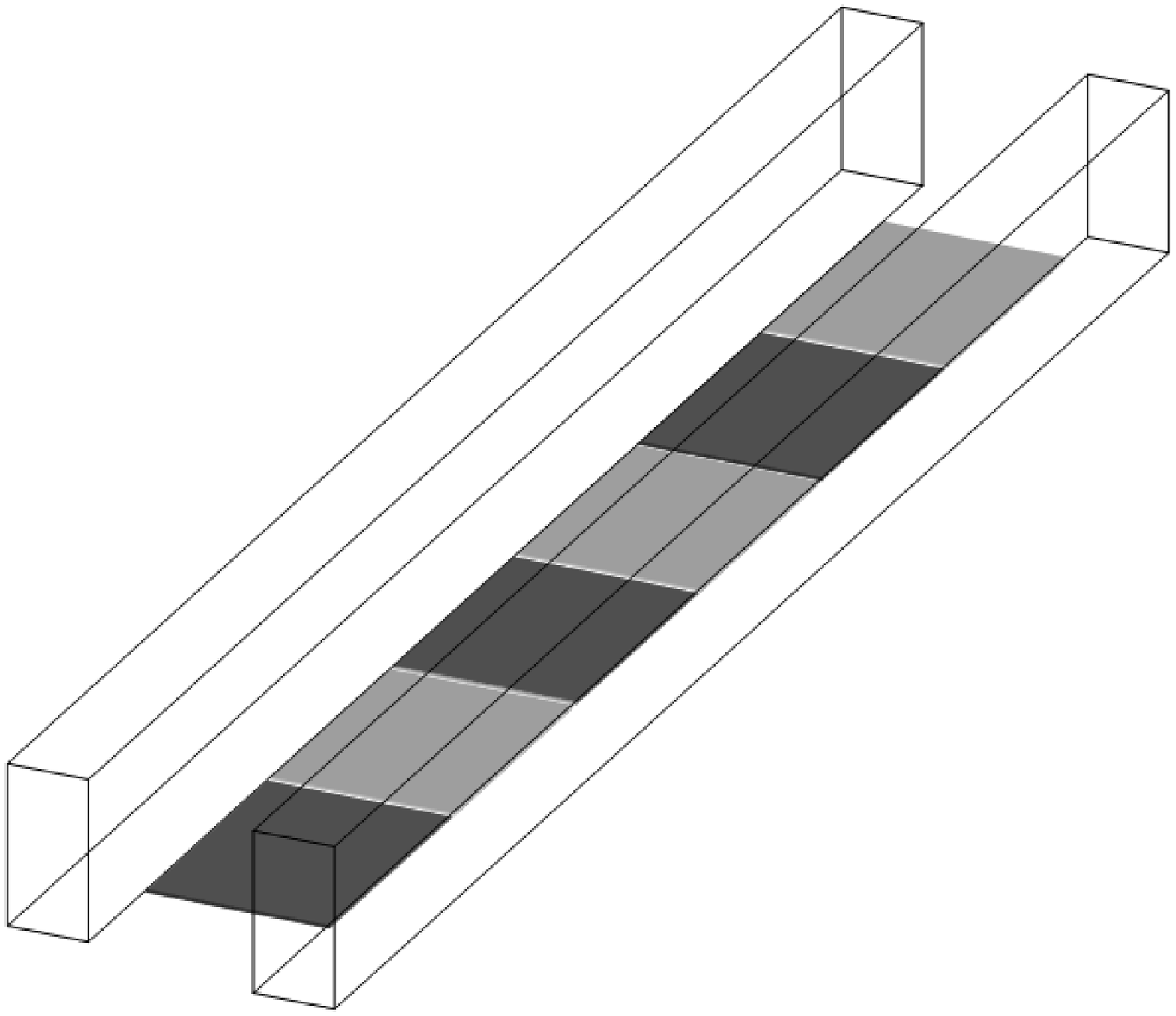}\includegraphics{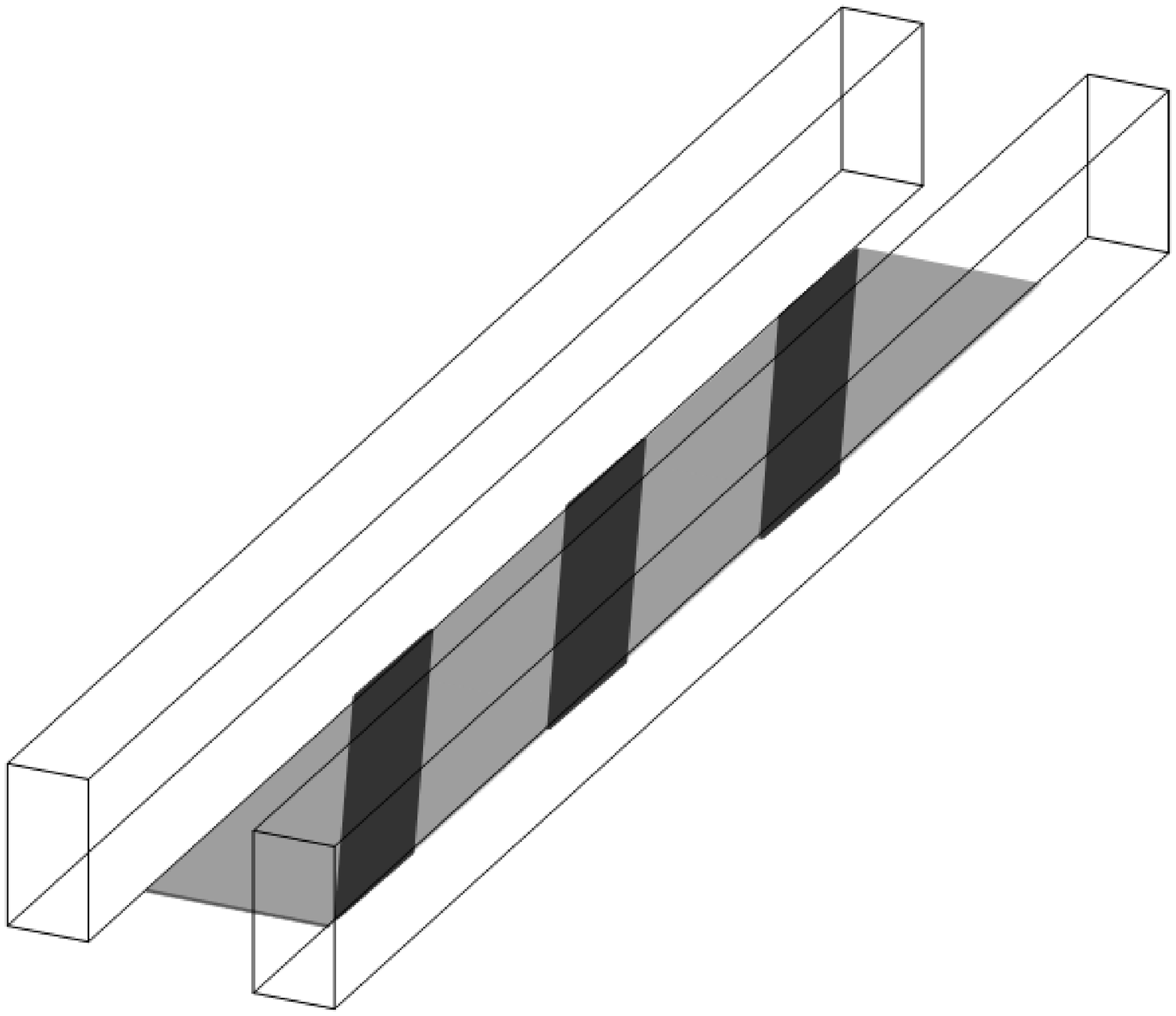}}
\caption{\label{mixers} Suggested designs for mixing devices. The light regions would be coated
in such a way as to induce a large slip length (say with a superhydrophobic coating), while the
dark regions would be coated to induce a small slip length or no slip (say with a superhydrophilic
coating). More complicated patterns may enhance the mixing, provided the patterns are on a length
scale comparable to the channel width.}
\end{figure}

In any case our calculations here have demonstrated that in an incompressible Newtonian fluid, changes in slip length can induce flow transverse to the walls in a nano or microfluidic channel. Further our calculations suggest that these transverse flows are maximised if the patterning of slip takes place on a wavelength $\lambda \sim w$. Thus it certainly
appears that patterned slip could be used to induce mixing in the same way as patterned topography (e.g. such as the asymmetric herringbone pattern studied in Ref~\cite{Stroock02}). Figure~\ref{mixers} suggests some possible patternings that could be used for mixing. However, we note that slip also changes the velocity profile in a channel (e.g. see figure~\ref{patterned-slip}). These changes in profile will no doubt alter the effect of dispersion on mixing in a channel. While our perturbative approach is not valid in the limit of large variations in slip-length which would be most effective for mixing, we would expect the flows to be qualitatively similar. Cottin-Bizonne et al \cite{Cottin-Bizonne03} have calculated effective slip-lengths in a half-plane geometry for flows over no-slip and partial or full slip patterned regions, although they have not examined how this alters the velocity profiles.     

We also note that surfaces with switchable wettability have recently been demonstrated \cite{Lahann03}. This switchability suggests the interesting prospect of a slip length which is time and space dependent i.e. $\delta = \delta(x, t)$. The approach outlined in section 2 can easily be adapted to consider this situation. If we imagine a traveling wave variation in slip length $\delta ( \omega t + k x)$, then in a frame comoving with this wave, the effects on the flow will appear similar to those of speeding up (or slowing down) the flow. Thus the response of the fluid to a rapidly changing time-dependent slip will lag these changes in slip (as the fluid response does for the fast moving flow in figure~\ref{ulag-plot}). We will consider this problem in more detail in further work.

\section{Conclusion}

We have considered the flow of a Newtonian fluid in a channel with spatially varying surface
properties. Using a pertubative approach we derived equations that describe flow in a channel
with patterned variations in slip length. We also examined flows in a channel with varying 
wettability using molecular dynamics. The simulations demonstrated that the variations in wettability
induce variations in slip. Good qualitative agreement was found between the molecular dynamics simulations 
and the approximate analysis of the Navier-Stokes equations. 

\begin{acknowledgments}
The authors wish to acknowledge funding from the MacDiarmid Institute for Advanced Materials
and Nanotechnology. The authors would also like to thank Cecile Cottin-Bizonne and Catherine 
Barentin for providing us with several useful references.
\end{acknowledgments}


\thispagestyle{empty}

\clearpage

\end{document}